\documentclass[%
 reprint,
 amsmath,amssymb,
 aps,
]{revtex4-2}
\usepackage{graphicx}
\usepackage{dcolumn}
\usepackage{bm}
\usepackage{hyperref}
\usepackage{xcolor}
\definecolor{Fire}{RGB}{234,92,12}
\definecolor{Water}{RGB}{95, 130, 193}
\definecolor{Sky}{RGB}{152, 197, 221}
\definecolor{Purple}{RGB}{103, 82, 158}
\hypersetup{
    colorlinks=true,
    linkcolor=Water,
    citecolor=Water,
    urlcolor=Purple,
    pdftitle={Quantum statistical mechanics on a digital quantum computer},
    pdfkeywords={Trapped Ions} {DQC1} {Kernel Polynomial Method} {Quantum Simulation} {Density of States},
    pdfnewwindow=true,
    }

\usepackage{xcolor}

\makeatletter
\newsavebox{\@brx}
\newcommand{\llangle}[1][]{\savebox{\@brx}{\(\m@th{#1\langle}\)}%
  \mathopen{\copy\@brx\kern-0.5\wd\@brx\usebox{\@brx}}}
\newcommand{\rrangle}[1][]{\savebox{\@brx}{\(\m@th{#1\rangle}\)}%
  \mathclose{\copy\@brx\kern-0.5\wd\@brx\usebox{\@brx}}}
\makeatother

\usepackage{dsfont}     
\usepackage{mathtools}
\usepackage{braket}

\newcommand{\abs}[1]{\left\lvert#1\right\rvert}     
\newcommand{\norm}[1]{\left\lVert#1\right\rVert}    
\newcommand{\tr}{\text{Tr}}    
\newcommand{\h}{\hat{H}}    
\newcommand{\hilb}{\mathcal{H}}    
\newcommand{\ketbra}[2]{\lvert#1\rangle\!\langle#2\lvert}
\newcommand{\id}{\hat{I}}    
\usepackage{xspace} 

\newcommand*{\figref}[2][]{\hyperref[{fig:#2}]{Fig.~\ref*{fig:#2}\ifx\\#1\\\else.($#1$)\fi}}

\newcommand*{\appref}[2][]{\hyperref[{app:#2}]{Appendix~\ref*{app:#2}}}

\renewcommand*{\eqref}[2][]{\hyperref[{#2}]{Eq.~(\ref*{#2})}}

\newcommand*{\secref}[2][]{\hyperref[{#2}]{Section~\ref*{#2}}}

\begin{document}

\preprint{}

\title{Calculating the many-body density of states on a digital quantum computer}

\author{Alessandro Summer}
 \email{summera@tcd.ie}
\affiliation{School of Physics, Trinity College Dublin, Dublin 2, Ireland}
\author{Cecilia Chiaracane}
 \email{chiaracc@tcd.ie}
\affiliation{School of Physics, Trinity College Dublin, Dublin 2, Ireland}
\author{Mark T. Mitchison}
 \email{mark.mitchison@tcd.ie}
\affiliation{School of Physics, Trinity College Dublin, Dublin 2, Ireland}
\author{John Goold}
 \email{gooldj@tcd.ie}
\affiliation{School of Physics, Trinity College Dublin, Dublin 2, Ireland}


\date{\today}

\begin{abstract}
Quantum statistical mechanics allows us to extract thermodynamic information from a microscopic description of a many-body system. A key step is the calculation of the density of states, from which the partition function and all finite-temperature equilibrium thermodynamic quantities can be calculated. In this work, we devise and implement a quantum algorithm to perform an estimation of the density of states on a digital quantum computer which is inspired by the kernel polynomial method. Classically, the kernel polynomial method allows to sample spectral functions via a Chebyshev polynomial expansion. Our algorithm computes moments of the expansion on quantum hardware using a combination of random state preparation for stochastic trace evaluation and a controlled unitary operator. We use our algorithm to estimate the density of states of a non-integrable Hamiltonian on the Quantinuum H1-1 trapped ion chip for a controlled register of 18 qubits. This not only represents a state-of-the-art calculation of thermal properties of a many-body system on quantum hardware, but also exploits the controlled unitary evolution of a many-qubit register on an unprecedented scale.
\begin{description}
\item[Tags]
Quantum Simulation, Kernel Polynomial Method, DQC1, Density of States, Trapped Ions
\end{description}
\end{abstract}

\maketitle

\section{Introduction}
The idea of using one quantum system to efficiently simulate another one was the vision of Feynman over 40 years ago~\cite{feynman1982simulating}. This paradigm is known as quantum simulation~\cite{lloyd1997,Georgescu_14,Gerace_20,daley2022practical} and is expected to be one of the first real applications of the current generation of quantum computers~\cite{Preskill2018quantumcomputingin}. In particular, recent progress has been made in simulating the dynamics of strongly correlated many-body systems on current devices~\cite{zhukov2018algorithmic,CerveraLierta2018exactisingmodel,francis2020quantum,smith2019simulating,keenan2022evidence}, albeit with systems which are still too small to compete with calculations on classical super-computing architectures. The hope is that the achievable system sizes will eventually become large enough to surpass what is classically possible. 

In terms of using quantum simulators to extract eigen energies of many-body systems, early ideas include algorithms based on quantum Fourier transform~\cite{,PhysRevLett.83.5162} such as quantum phase estimation~\cite{kitaev1997quantum, cleveQuantumAlgorithmsRevisited1998, abramsQuantumAlgorithmProvidingExponential1999} and adiabatic state preparation~\cite{aspuru2005simulated}. The development of algorithms for the extraction of ground-state energies is central for the promise of being able to perform quantum chemistry and materials simulations on quantum computers~\cite{kassal2011simulating,hastings2015improving,cao2019quantum,de2021materials} and ground state energy calculation is a target of many variational quantum algorithms~\cite{cerezo2021variational,RevModPhys.94.015004}. Results for finite temperature and excited states are more scarce. However recent proposals to measure finite temperature expectation values on hardware include sampling~\cite{temme2011quantum,chowdhury2017quantum,PhysRevA.102.022622} and imaginary time evolution~\cite{motta2020determining} and more recently algorithms which may have potential for computing micro-canonical expectation values were proposed in~\cite{PRXQuantum.2.020321}.

The more general idea of using quantum computers to do statistical mechanics is a topic which is gaining traction~\cite{PRXQuantum.2.020321,schuckert2022probing}. In this work we focus on developing an algorithm that gives a coarse-grained estimate of the density of states (DOS) based on the classical kernel polynomial method (KPM)~\cite{weisseKernelPolynomialMethod2006}. The KPM provides a reconstruction of a spectral function by means of a Chebyshev polynomial expansion, weighted by suitable kernels to damp the Gibbs oscillations that occur due to finite series truncation. Chebyshev moments are computed iteratively by applying functions of the Hamiltonian on some initial state. This step is a challenge to implement on quantum hardware.

Block encoding of a Hamiltonian is deeply connected with the Chebyshev polynomials~\cite{childsQuantumAlgorithmSystems2017}, implementing the Hamiltonian as a quantum walk as exploited in the context of the KPM in~\cite{rallQuantumAlgorithmsForEstimating2020} and more generally to estimate physical properties in~\cite{roggeroSpectralDensityEstimation2020, rajputHybridizedMethodsQuantumSimulation2022}. An alternative is to compute the Chebyshev moments iteratively in a variational quantum algorithm~\cite{jensenNearTermQuantumAlgorithm} or otherwise overcoming the problem of implementing the Chebyshev polynomials using suitably defined Fourier ones~\cite{wang2022quantum, hartseFasterSpectralDensityCalculation2022}.

In this work we devise a hybrid algorithm which uses a combination of pseudo-random state preparation, Hadamard test and Suzuki-Trotter (ST) decomposition~\cite{suzukiGeneralTheoryFractalPath1991} to evaluate Chebyshev moments. These moments are then used in the standard KPM expansion. We use an arc-cosine approximation of the Hamiltonian to implement Chebyshev polynomials from standard ST decomposition and implement our algorithm on the Quantinuum H1-1 trapped ion quantum simulator~\cite{quantinuum}. We were able to approximate the DOS of a non-integrable spin chain for up to 18 qubits using a single ancillary qubit. Our simulations represent one of the first explorations of the use of near term quantum computers for calculations in statistical mechanics. 

In \secref{sec:kpm} we introduce the KPM method and discuss its use in the context of statistical mechanics. We discuss how to compute the DOS and how a pseudo random state can be used for stochastic trace estimation. In \secref{sec:kpmquantum} we explain the quantum algorithm for extracting Chebyshev polynomials and discuss the subroutines for random state preparation and implementing the arc-cosine approximation of the Hamiltonian. In \secref{sec:results} we then introduce the model we simulate on hardware and the corresponding gate decomposition used to implement the controlled unitary. We display our results for estimations of DOS computed using our hybrid algorithm for systems sizes of $12$ and $18$ qubits.

\section{Classical Kernel Polynomial Method for the density of states}
\label{sec:kpm}

\subsection{Density of states}

In this section we give an overview of the classical KPM and discuss how it is used to calculate the DOS~\cite{weisseKernelPolynomialMethod2006}. For a system of $L$ qubits with Hamiltonian $\hat{H}$, the DOS is defined as
\begin{equation}
\label{dos}
    g(E) = \frac{1}{2^L}\sum_{k=0}^{2^L-1} \delta (E-E_k),
\end{equation}
where we denote the energy eigenvalues by $E_k$ and the corresponding eigenvectors by $\ket{k}$, i.e. $\hat{H}\ket{k} = E_k\ket{k}$. The DOS gives access to all thermodynamic properties: in particular, the canonical partition function can be evaluated as
\begin{equation}
    \mathcal{Z}(\beta)=\int E e^{-\beta E} g(E) dE.
\end{equation}
Differentiation of $\mathcal{Z}$ yields any desired thermodynamic quantity: for example, the energy 
\begin{equation}
   E(\beta)=-\frac{\partial(\log\mathcal{Z}(\beta))}{\partial\beta}, 
\end{equation}
 and the entropy $S(\beta)=\beta(E(\beta)-F(\beta))$ where $F(\beta)=-\beta^{-1}\log(\mathcal{Z(\beta)})$ is the free energy. 
 
 In order to extract the DOS from a system of size $L$ one would typically need exact diagonalization, which requires memory resources scaling as $\mathcal{O}(2^{3L})$. In contrast, the KPM described in the following section is able to approximate with memory scaling only as $\mathcal{O}(2^{L})$ (combined with stochastic evaluation of the trace, analysed in \secref{sec:stoch_trace}). 

\subsection{Kernel polynomial method}

Consider some function $f(x)$ defined on the interval $x\in [-1,1]$. The KPM provides an optimal approximation of this function by a finite series of $M$ Chebyshev polynomials. Mathematically, it is defined as
\begin{equation}
    \label{eq:kpm_of_f}
    f_\text{KPM}(x) = \frac{1}{\pi\sqrt{1-x^2}}\sum_{m=0}^M \gamma_m^M \mu_m T_m(x),
\end{equation}
where $\gamma_m^M$ are the kernel coefficients used to damp Gibbs oscillations, $T_m$ are the Chebyshev polynomials, and $\mu_m$ are the Chebyshev moments. 
While a more detailed discussion can be found in \appref{chebyshev_polynomials}, the polynomials are generally defined as 
\begin{equation}\label{eq:cheb_as_cosarc}
    T_m(x)=\cos(m\arccos(x))\quad \text{with}\  m\in \mathbb{N}_0.
\end{equation}
For example, $T_0(x) = 1$, $T_1(x) = x$, and all higher Chebyshev polynomials obey the recursion relation
\begin{equation}
    \label{Chebyshev_recursion}
    T_{m}(x) = 2xT_{m-1}(x) - T_{m-2}(x).
\end{equation}
The KPM expansion~\eqref{eq:kpm_of_f} is thus reconstructed by computing the corresponding Chebyshev moments
\begin{equation}\label{eq:cheb_mom_f}
    \mu_m(x)=\int_{-1}^1 f(x) T_m(x) \mathrm{d}x.
\end{equation}

The KPM can be adapted to estimate general spectral functions involving a given quantum mechanical Hamiltonian. In this work, we will be interested specifically in using it to get an estimate for the DOS. We note that recently the KPM has been implemented using tensor network techniques~\cite{PhysRevB.90.115124}, and was used to study thermalisation~\cite{yang2020probing} as well as to extract the DOS of simple lattice gauge theories~\cite{PhysRevD.104.014514}. One clear advantage of the KPM is that it does not suffer from the sign problem that is synonymous with Monte-Carlo simulations. 

\subsection{Chebyshev moments of the DOS}

Let us discuss the key steps in computing the KPM approximation of the DOS, $g(E)$. Since the domain of Chebyshev polynomials is $[-1,1]$, $\h$ must have a spectral norm $\vert\vert\h\vert\vert\le 1$. If not, $\h$ can be normalised as
\begin{equation}
\label{eq:normilised_H}
    \h\mapsto \frac{\h-a}{b},
\end{equation}
with
\begin{equation}
\label{eq:normalisation}
     a\coloneqq\frac{E_\text{max}+E_\text{min}}{2}\quad\text{and } \quad b\coloneqq\frac{E_\text{max}-E_\text{min}}{2-\varepsilon}
\end{equation}
where $E_\text{max}$ and $E_\text{min}$ are the largest and smallest eigenvalues of $\h$. Additionally, a small cutoff $\varepsilon$ is introduced to avoid stability issues that can arise close to the boundaries of the spectrum. After this rescaling is performed, the expression of the moments $\mu_m$ becomes
\begin{equation}
\begin{split}\label{eq:mu_def}
    \mu_m \coloneqq& \int_{-1}^1 g(E)T_m(E)\mathrm{d}E\\
    =&\int_{-1}^1 \frac{1}{2^L}\sum_{k=0}^{2^L-1} \delta (E-E_k)  T_m(E) \mathrm{dE} \\ 
    =& \frac{1}{2^L} \sum_{k} T_m(E_k)= \frac{1}{2^L} \sum_k
    \bra{k} T_m(\h)\ket{k}\\ 
    =&\frac{1}{2^L}\tr[T_m(\h)].
\end{split}
\end{equation}

\subsection{Stochastic trace evaluation}
\label{sec:stoch_trace}
The first task in the extraction of Chebyshev moments of the KPM is the efficient estimation of the trace of an operator $\hat{X}$ acting on $L$ qubits, as in \eqref{eq:mu_def}. For example computational complexity of determining an element on the main diagonal, $X_{ii}\coloneqq\bra{i}\hat{X}\ket{i}$, is $\mathcal{O}(2^L)$, and therefore determining all elements on the main diagonal requires $\mathcal{O}(2^{2L})$ operations~\cite{jinRandomStateTechnology2021}.  
A common alternative method to this computational procedure is stochastic trace estimation. The main idea is to estimate $\hat{X}$ on a set of randomly chosen states.
Let $\{\ket{r}\}$ be a set of $R$ random states on $L$ qubits:
\begin{equation}
    \ket{r} = \sum_{i=0}^{2^L-1} c_{ri} \ket{i}\quad\text{with }c_{ri}\in\ \mathbb{C},
\end{equation}
so that the stochastic estimate will be
\begin{equation}
    \Theta =\frac{1}{R} \sum_{r} \bra{r} \hat{X} \ket{r}.
\end{equation}
If $\forall r$ the statistical average is $0$, i.e. 
\begin{equation}
\label{eq:rnd_cond1}
    \llangle c_{ri}\rrangle = 0,
\end{equation} (where the $\llangle \cdot \rrangle$ average is over $i$), and if 
\begin{equation}
\label{eq:rnd_cond2}
\begin{split}
    \llangle c_{ri} c_{r'j}\rrangle &= 0,\\
    \llangle c_{ri}^* c_{r'j}\rrangle &= \delta_{r r'}\delta_{ij},
\end{split}
\end{equation}
then the variance in the estimate of $\tr[\hat{X}]$ would be
\begin{equation}\label{eq:fourth_mom_requirement}
    \big(\delta \Theta \big)^2 = \dfrac{1}{R}\left(\tr[\hat{X}^2] + (\llangle \abs{c_{ri}}^4 \rrangle -2)\sum_{j=0}^{2^L-1} X_{jj}^2\right)
\end{equation}
(for more details see the overview in~\cite{weisseKernelPolynomialMethod2006}). Therefore, the relative error will scale as $\mathcal{O}\big(1/\sqrt{R2^L}\big)$, which means that fewer random states will be needed to achieve a fixed level of precision as the size of the system increases.
Note that if the coefficients of the random states are distributed as a Gaussian, then the variance of $\Theta$ only depends on $\tr[\hat{X}^2]$. 

The stochastic trace is used in conjunction with the recursion relation~\eqref{Chebyshev_recursion} to estimate the Chebyshev moments.
Starting with a random state $\ket{r} = \ket{r_0}$, one can define
\begin{equation}
\ket{r_1} = \h \ket{r_0},
\end{equation}
and recursively generate a series of $M$ vectors
\begin{equation}
\ket{r_m} = 2\h \ket{r_{m-1}} - \ket{r_{m-2}} = T_m(\hat{H})\ket{r}.
\end{equation}
The moments can then be obtained from the overlap $\braket{r_m|r} = \braket{r|T_m(\hat{H})|r}$. 

One potential issue with computing the moments in this way is that errors make the precision of $m$-th moment dependent on errors from the evaluation of the previous ones. As mentioned earlier, the computation of the $\mu_m$ is the major bottleneck of the classical KPM method, limiting the size of the systems it can be applied to. In the next section we describe our algorithm to compute the $m$-th Chebyshev moment on a quantum computer. 

\section{Kernel polynomial method hybrid quantum algorithm}
\label{sec:kpmquantum}

\begin{figure}[!ht]
\centering{\includegraphics[width=1.0\columnwidth]{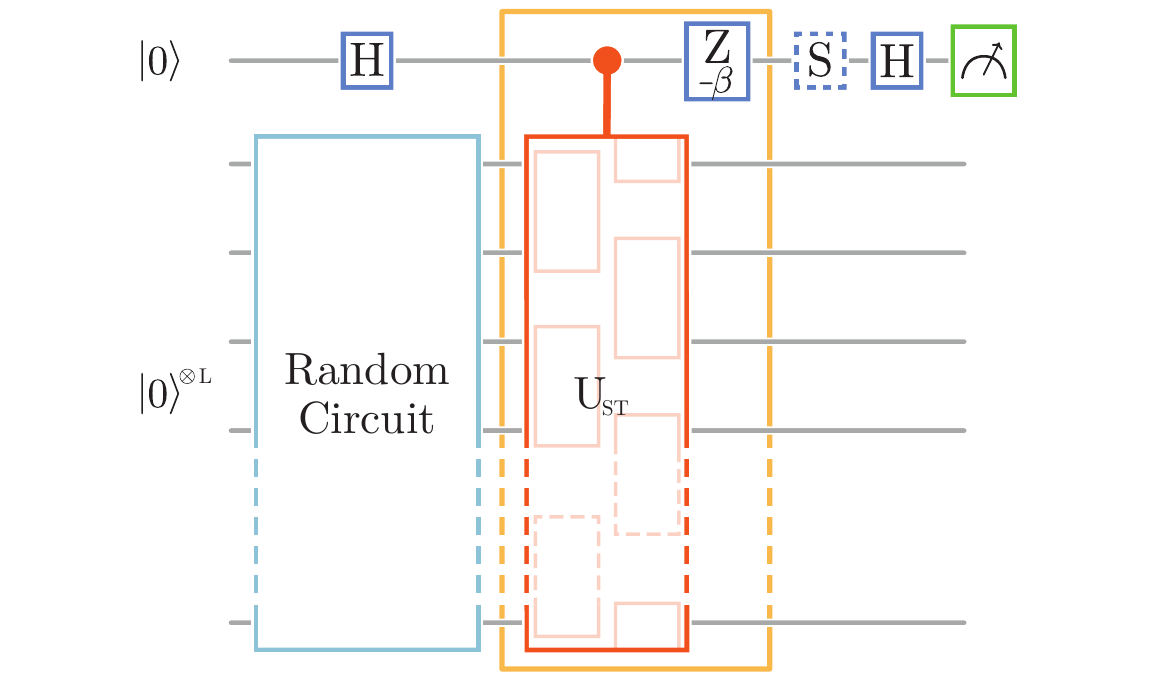}
\vspace{-.75cm}
\caption{The proposed circuit to compute the Chebyshev moments on the quantum computer is shown. A pseudo-random state is generated by means of a random circuit on a register of $L$ qubits and an ancillary qubit is prepared in the $\ket{+}$ state with a Hadamard operation.  A  unitary operation $e^{im\h_K}$ is then performed on the register and controlled on the ancillary qubit (yellow rectangular). This is pursued through a controlled unitary and a rotation $Z$, with $\beta$ the coefficient of the identity in $\h_K$. The subsequent gates and measurements on the controlled qubit yield $\tr[\cos(m\h_K)]$ and $\tr[\sin(m\h_K)]$, with $\hat{H}_K$ connected to the arc-cosine expansion defined by \eqref{eq:arcc} which can be then connected to the Chebyshev polynomials of the Hamiltonian of order $m$.}
\label{fig:final_circuit}
}\end{figure}

\figref{final_circuit} shows the circuit we use to compute Chebyshev moments on quantum hardware. Our proposal is reminiscent of the DQC1 protocol~\cite{knillPowerOneBit1998}, which is a sub-universal computational paradigm that requires $L+1$ qubits, an ancillary one (that will act as a control) and $L$ qubits initialised in a maximally mixed state. 
The circuit consists of a Hadamard test where a controlled unitary is applied to $L$ qubits and then the control qubit is measured. We simulate the maximally mixed state by a randomisation procedure.
We now discuss the two main parts of the circuit in detail. \secref{subsec:randomiser} analyses the random circuit we use to mimic the identity register and \secref{subsec:implenting_cheb} the application of the controlled unitary for the extraction of Chebyshev moments.

\subsection{Stochastic trace evaluation via state randomisation}
\label{subsec:randomiser}
\begin{figure}[!ht]
\centering{\includegraphics[width=1.\columnwidth]{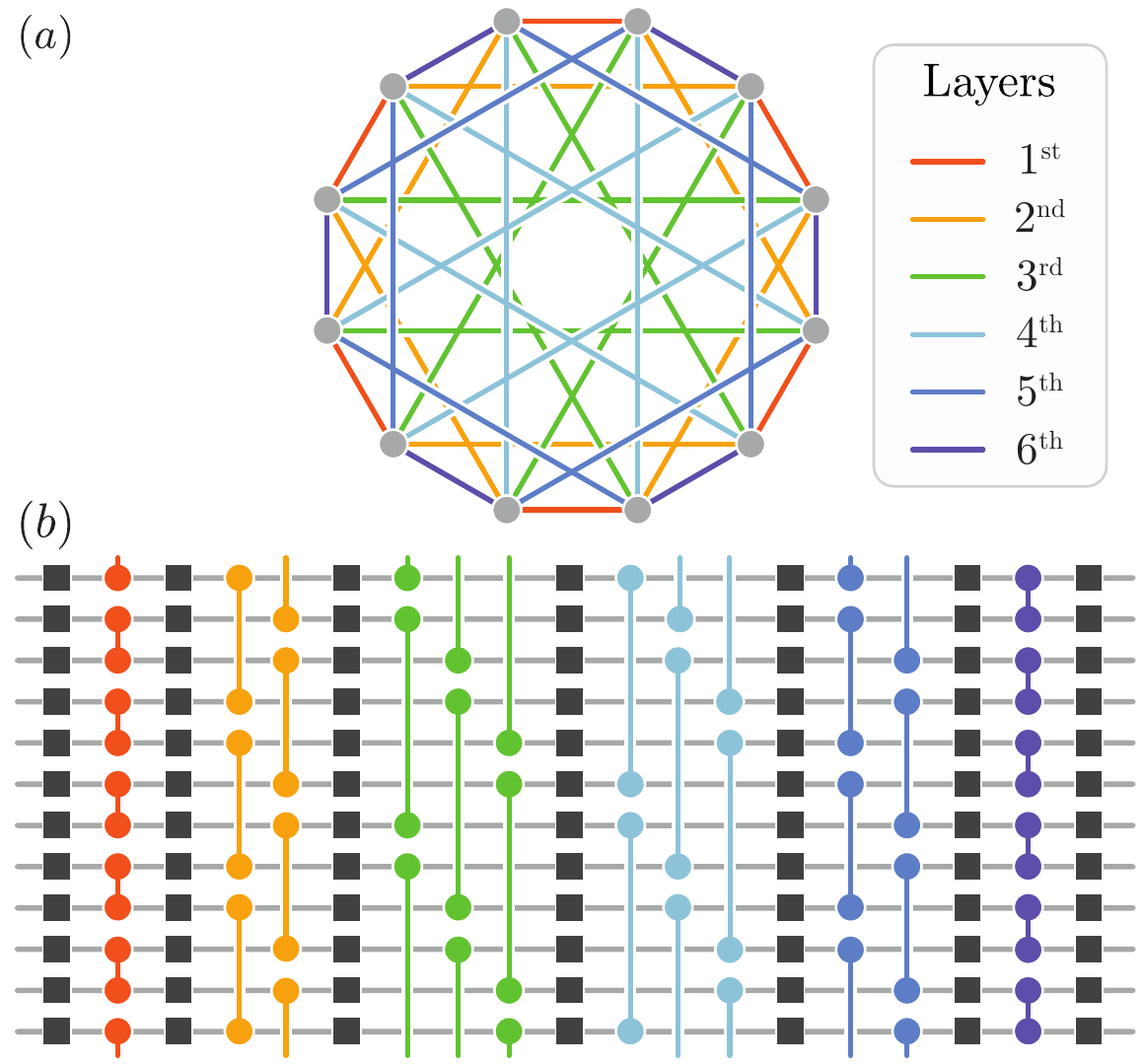}}
\vspace{-.75cm}
\caption{Shown in this figure is the 6-layer randomised circuit used for trace evaluation in the $L=12$ (the blue box on L-1 qubit register in \figref{final_circuit}). ($a$) A representation that illustrates the connectivity of the circuit, the grey dots represent the qubits, and the colored lines are the 2-qubit gates where they are numbered according to their application. Note that this particular structure is chosen to exploit the all to all connectivity of the Quantinuum H1-1 device~\cite{quantinuum}. ($b$) Gate decomposition of the random circuit. The colored lines represent $\hat{ZZ}(\pi/2)$ gates, and the dark gray squares are the single-qubit rotations, randomly selected from $\hat{X}(\pi/2)$, $\hat{Y}(\pi/4)$ and $\hat{Z}(\pi/2)$. The latter two will be removed by the compilation since they commute with $\hat{ZZ}(\pi/2)$ and anti-commute with the other operators. Therefore, their effect is to add two more types of single-qubit rotations.}
\label{fig:random_circuit}
\vspace{-.25cm}
\end{figure}

The generation of random states on quantum computers has garnered considerable recent attention recently due to their important role in benchmarking~\cite{boixoCharacterizingQuantumSupremacy2018}. However, 
producing uniformly random states that sample the Haar distribution is inefficient in the sense that the number of gates required scales exponentially with the register size~\cite{nielsenQuantumComputationQuantumInformation2002, emersonPseudoRandomUnitaryOperators2003}.
T-designs are a type of circuit that replicate moments of the Haar distribution up to order T or lower~\cite{ambainisQuantumTDesigns2007}. They offer some improvement in terms of gate efficiency, with the number of gates required scaling polynomially with the number of qubits. Nonetheless, T-designs are still costly and may be overly random for specific purposes.
Here we take an alternative approach based on pseudorandom state generation~\cite{emersonPseudoRandomUnitaryOperators2003}. This involves the generation of states that do not uniformly sample the Haar distribution~\footnote{In some particular cases, it has been proven that the pseudorandom states after a certain depth become T-designs~\cite{harrowApproximateUnitaryDesigns2018}}, but still
possess the desired properties such as Eqs.~\hyperlink{eq:rnd_cond1}{(\ref{eq:rnd_cond1})} and \hyperlink{eq:rnd_cond2}{(\ref{eq:rnd_cond2})}. This method has been shown to generate states that are sufficiently random in an efficient way.

Using pseudorandom states to stochastically evaluate the trace has been proposed in various papers~\cite{wang2022quantum, richterSimulatingHydrodynamicsNoisy2021a, sekiEnergyFilteredRandomPhaseStates2022, gotoMatrixProductStateApproach2021, gotoEvaluatingThermalExpectationValues2023}, and recently used on quantum hardware to extract high-temperature transport exponents~\cite{keenan2022evidence}. In order to generate pseudo-random states, one can use a circuit composed of alternating layers of 2-qubit gates and layers with random single qubit rotations, as suggested by Ref.~\cite{emersonPseudoRandomUnitaryOperators2003} and adopted in Refs.~\cite{boixoCharacterizingQuantumSupremacy2018, richterSimulatingHydrodynamicsNoisy2021a, keenan2022evidence} with small variations. 
This random composition has become a widely accepted method for generating this type of random state. In Ref.~\cite{richterSimulatingHydrodynamicsNoisy2021a} the random state is aimed at generation on a quantum computer where the qubits are connected in a ring geometry. Layers of 2-qubit gates connecting even-odd and odd-even qubits are alternated and in between them, there are layers of single-qubit gates randomly chosen among $\{\hat{X}(\pi/2), \hat{Y}(\pi/2), \hat{Z}(\pi/4)\}$ so that the same rotation is not applied to the same qubit sequentially.

In our case, we focus on a variation of this procedure that takes advantage of the all-to-all connectivity and of the specific set of elementary gates that can be implemented directly on the Quantinuum H1-1 trapped-ion-based quantum computer (see \figref{random_circuit}). 
The gate set of the device Quantinuum H1-1 device includes
\begin{equation}
\label{eq:gates_quantinuum}
    \begin{split}
        \hat{ZZ}(\theta) &= e^{-i\theta/2 \hat{Z}\otimes \hat{Z}},\\
        \hat{Z}(\theta) &= e^{-i\theta/2 \hat{Z}},\\
        \hat{U}_{1q}(\theta, \phi) &= e^{-i\theta/2 (\cos(\phi) \hat{X} + \sin(\phi) \hat{Y})}.\\
    \end{split}
\end{equation}
The device that we will use also supports parallelisation (using Quantum charge-coupled device (QCCD) architecture with five parallel gate zones~\cite{pinoDemonstrationTrappedIon2021}).
Inspired by existing techniques to create shallow randomisers, we change the single qubit rotations to be chosen from  $\{\hat{X}(\pi/2), \hat{Y}(\pi/4), \hat{Z}(\pi/2)\}$.  Note that the former two rotations can be applied as a single $\hat{U}_{1q}(\theta, \phi)$ gate, while the latter can be implemented virtually~\footnote{Anyway we can noticed that $\hat{Z}$ commutes or anti-commutes with all the gates so it could be implemented  virtually independently from the device}. 
The 2-qubit gate will now be a $\hat{ZZ}(\pi/2)$ with a different connectivity. Each $\hat{ZZ}(\pi/2)$ will connect the qubits $2i$ and $(2i+p)\mod L$ with $i=0,\dots, L/2$ and $p$ an odd number called a \textit{jump}. The jumps are chosen so that each qubit is narrowly connected to the other. For instance, as shown in \figref{random_circuit}, the jumps of the first half of the layers can be chosen with $p_\ell=-(-1)^\ell (2s\ell+1)$ with $s\in\mathbb{N}_0$ and $\ell$ the index of the layer, and with $p_\ell=-p_{-\ell}$ for the second half of the layers. In particular, \figref{random_circuit} shows the random compiler with $L=12$ and $s=1$. Notice that $s=0$ reproduces the same pattern of 2-qubit gates as in~\cite{richterSimulatingHydrodynamicsNoisy2021a}. 

Following Ref.~\cite{richterSimulatingHydrodynamicsNoisy2021a}, we quantify the randomising effect of our circuit by checking how well the half-system entanglement entropy converges to the Page value~\cite{pageAverageEntropy1993}, which is the entanglement entropy for a typical Haar-random state. The von Neumann entanglement entropy of a state $\rho_r=\ketbra{r}{r}$ on a space divided in subspaces $\hilb_a$ and $\hilb_b$ is:
\begin{equation}
    S(\rho_{r(a)}) = \tr[\rho_{r(a)}\ln \rho_{r(a)}]
\end{equation}
with $\rho_{r(a)} = \tr_{(b)}[\ketbra{r}{r}]$.
As shown in~\cite{pageAverageEntropy1993}, this value converges to the Page value $\log(\dim\hilb_a)-\dim\hilb_a/(2\dim\hilb_b)$ for random pure states. For $\dim\hilb_a=\dim\hilb_b$ the it becomes $\log(2^{L/2})-1/2.$

\begin{figure} 
\centering{\includegraphics[width=1.\columnwidth]{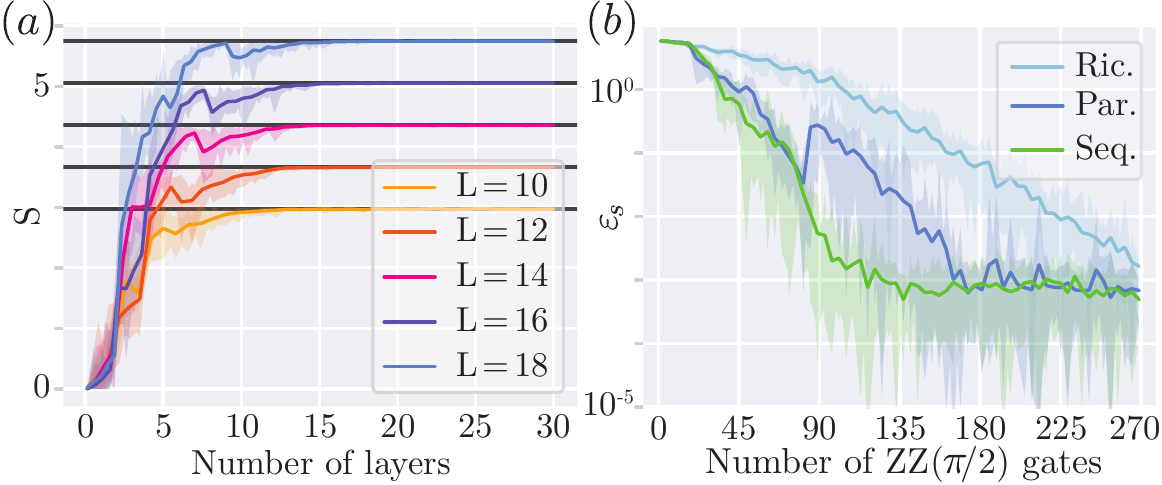}}
\vspace{-.75cm}
\caption{The convergence of different half-chain entanglement entropies to the expected value,  for $L=10, 12, 14, 16$ and $18$ averaged over $20$ different random states. The shadows are drawn by the values of the individual random states. ($a$) The half-chain von Neumann entropy $S$ is shown as function of random circuit depth and shows the saturation to the Page value~\cite{pageAverageEntropy1993} (dark grey lines) for different system size for the circuit used in this work (Par.). ($b$) Comparison of the relative error of the von Neumann half-chain entropy $\varepsilon_s$ at $L=18$ of the three different random circuits: the one proposed in Richter \textit{et al.}~\cite{richterSimulatingHydrodynamicsNoisy2021a} (Ric.), the one used in this work (Par.) (see Fig. 2) and one where the 2-qubit gates are applied sequentially (Seq.). All three circuits lead to fast convergence to a maximally bi-partite entangled state. We note that the relative error in Seq. seems to be most favorable but we have used Par. on the hardware here because less layers are required to get a good estimate of the trace.\label{fig:entropy_convergence}}
\vspace{-.25cm}
\end{figure}
 
\begin{figure}
\centering{\includegraphics[width=1.0\columnwidth]{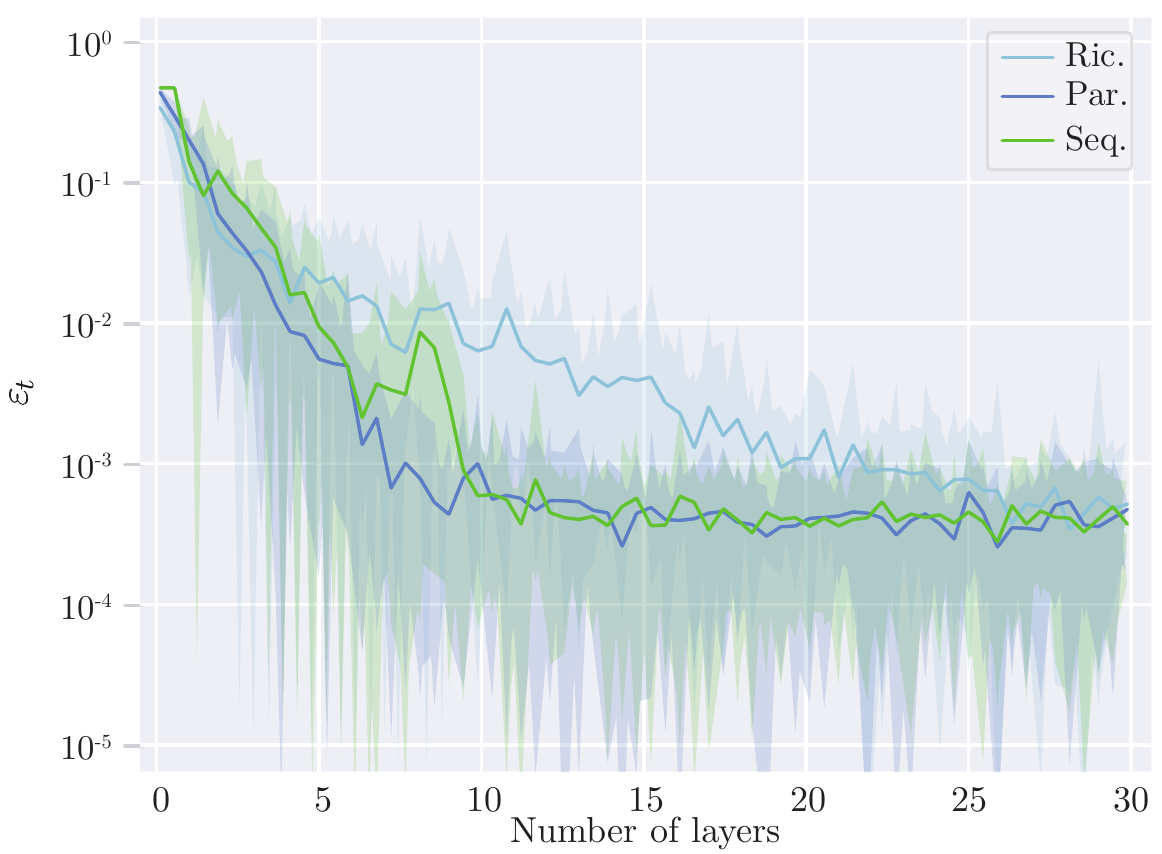}
\vspace{-.75cm}
\caption{
   Convergence of relative error $\varepsilon_t$ of the stochastic trace to the trace of the Hamiltonian defined by \eqref{eq:xyz_zz} with $L=18$ for the three random circuits used to generate \figref{entropy_convergence}. The results are averaged over $20$ different random states from which the shadows are obtained. 
} \label{fig:stoc_trace}
}\end{figure}

In \figref{entropy_convergence}, we present a classical simulation comparing the convergence rates to the Page value of three different approaches: Ric., which is the approach suggested by Richter et al.~\cite{richterSimulatingHydrodynamicsNoisy2021a}; Par., the approach used in this work; and Seq., a variation of Par. where the 2-qubit gates are applied as a chain. In this latter method, the layers of 2-qubit gates are composed of a sequence of gates,  each of these has support overlapping over half of the previous and half of the succeeding gate. Therefore, we will compare these three methods in terms of the number of 2-qubit gates used instead of the number of layers.
Although this last constraint limits the implementation on Quantinuum's hardware it is helpful in analysing the three different random compilers. 
In fact, while Par. is not the fastest at converging to the Page value, it performs best at estimating the trace, as shown in \figref{stoc_trace}.
In contrast, the states produced by Seq. converge to the Page value fastest, but they are not as effective for trace estimation as those built with Par. This suggests that the entanglement entropy alone is not sufficient to determine a suitable state for stochastic trace estimation. In fact, we have observed that, as the entropy converges to the Page value, the fourth moment of the coefficients $c_{ri}$ converges to 2 (as required by \eqref{eq:fourth_mom_requirement}). This suggests that the states become Gaussian-distributed at the same rate as they approach the Page value.

\subsection{Implementing the Chebyshev polynomials}
\label{subsec:implenting_cheb}
Simulating the unitary evolution operator or other functions of the Hamiltonian are central tasks in the field of 
quantum simulation.  In recent years there has been significant progress, with various approximations, like qDrift~\cite{campbellRandomCompilerForFastHamiltonian2019, chenConcentrationForRandomProductFormulas2021, faehrmannRandomizingMultiproductFormulas2022, berryTimeDependentHamiltonianSimulation2020}, LCU~\cite{childsHamiltonianSimulationUsingLinearCombinations2012}, Taylor expansion~\cite{berrySimulatingHamiltonianDynamics2015, berryHamiltonianSimulationWithNearlyOptimal2015} and qubitization~\cite{lowOptimalHamiltonianSimulation2017, lowHamiltonianSimulationQubitization2019}.
 The latter relies on the sequential application of two oracles: \textit{select} ($\hat{S}$) and \textit{prepare} ($\hat{P}$), defined as follows. Consider a normalised Hamiltonian that can be decomposed into a sum of unitary operators as
 \begin{equation}
 \label{eq:H_decomposition}
     \h = \sum_{p=1}^P {\omega_p} \h_p\qquad\text{with }\h_p^2=\id\text{ and }\omega_p\in\mathbb{R}.
 \end{equation}
 The select operator is defined as
 \begin{equation}
     \hat{S} \coloneqq \sum_{p=1}^P \ketbra{p}{p}_{(a)}\otimes \big(\h_{p}\big)_{(s)},
 \end{equation}
 where $(a)$ denotes an ancillary Hilbert space comprising $a$ qubits, where $a\geq  \lceil \log_2 P \rceil$ ~\cite{lowHamiltonianSimulationQubitization2019}.
 The prepare operator acts as
 \begin{equation}
     \hat{P} : \ket{0}_{(a)} \mapsto \sum_{p=1}^P\sqrt{\omega_p}\ket{p}_{(a)}\eqqcolon \ket{P}_{(a)}.
 \end{equation}
 These operators can be combined to obtain a block encoding of the Hamiltonian as
 \begin{equation}
      \bra{P}_{(a)} \hat{S} \ket{P}_{(a)} = \h.
 \end{equation}
 The idea behind qubitization is to exploit the block encoding of $\h$ in a quantum walk to generate any function of $\h$. 
 The \textit{walk operator} is defined as
 \begin{equation}
 \begin{split}
     \hat{W} \coloneqq & \underbrace{\big(2\ketbra{P}{P} - \id\big)_{(a)}}_{\hat{R}_{(a)}}\hat{S}\\
     =& \sum_{p,q} \big(2\sqrt{\omega_p\omega_q}\ketbra{q}{p} - \ketbra{p}{p}\big)_{(a)}\otimes \big(\h_p\big)_{(s)}
 \end{split}     
 \end{equation}
 where the operator $\hat{R}_{(a)}$ acts as a reflection about the $\ket{P}$ state. The walk operator can be decomposed as
 \begin{equation}
     \label{eq:walk_operator}
     \begin{split}
         \hat{W}=&\bigoplus_{k} \begin{pmatrix}
         E_k & \sqrt{I - \abs{E_k}^2}\\
         -\sqrt{I - \abs{E_k}^2} & E_k\\
     \end{pmatrix}\\
     =&\bigoplus_{k} e^{i\hat{Y}_{(k)} \arccos{E_k}}
     \end{split}
 \end{equation}
where $E_k$ are the eigenvalues (with respective eigenstates $\ket{k}$) of $\h$ and each term of the direct sum acts on the subspace $\hilb_k$ generated by $\ket{\phi_k}\coloneqq \ket{P}\ket{k}$ and its orthogonal state $\ket{\phi_k^\perp}\propto (\hat{S} - E_k \id)\ket{\phi_k}$. Likewise the $\hat{Y}_{(k)}$ operator will act as a Pauli $\hat{Y}$ operator on this subspace. \eqref{eq:walk_operator} is also useful to understand how $\hat{W}$ is isomorphic to
\begin{equation}
\label{eq:Y_arccosH}
    e^{i\hat{Y}\otimes \arccos \h}=\begin{pmatrix}
        \h & \sqrt{\id -\h^2}\\
        -\sqrt{\id -\h^2} & \h
    \end{pmatrix}
\end{equation}
that is the minimal block encoding of $\h$.
A fundamental feature of $\hat{W}$, on which the efficiency of qubitization is based, is that repeating it $m$ times and projecting it on $\ket{P}$ generates the Chebyshev polynomials:
\begin{equation}
    \bra{P}_{(a)} \hat{W}^m \ket{P}_{(a)} = \hat{T}_m(\h).
\end{equation}

Let us now consider the smallest decomposition of $\h$ in unitaries, where \eqref{eq:walk_operator} becomes easier to interpet. 
Given any Hamiltonian $\hat{H}$, there is a unitary operator $\hat{U}_H$ such that 
\begin{equation}
\label{eq:simple_decomp_of_h}
    \h=\frac12 (\hat{U}_H+\hat{U}_H^\dagger)
\end{equation}
where 
\begin{equation}
\label{eq:u_arccosH}
    \hat{U}_H \coloneqq \sum_k e^{i\arccos{E_k}}\ketbra{k}{k}=e^{i\arccos{\h}}
\end{equation}
It can be observed that $\hat{U}_H^2\neq \id$ as would seem to be required by~\eqref{eq:H_decomposition}; however, we notice that this condition can be relaxed by introducing a new walk operator $\hat{V}\coloneqq\hat{R}_{(a)}\hat{S}^\dagger$ and alternating it with $\hat{W}$.

As shown in detail in \appref{powers_of_h}, an alternative decomposition to \eqref{eq:simple_decomp_of_h} is 
\begin{equation}
    \h = \frac{1}{2i}\big(e^{i\arcsin{\h}}-e^{-i\arcsin{\h}}\big).
\end{equation}
For this decomposition of $\h$, the prepare operator simplifies to 
 \begin{equation}
     \hat{P} : \ket{0}_{(a)} \mapsto \frac{1}{\sqrt{2}}\ket{0}_{(a)}+ \frac{1}{\sqrt{2}}\ket{1}_{(a)}=\ket{+}_{(a)}
 \end{equation}
while the select operator becomes
\begin{equation}
    \hat{S}=e^{i\hat{Z}_{(a)}\otimes \arccos{\h}_{(s)}}=\begin{pmatrix}
        \hat{U}_H & 0 \\ 0 & \hat{U}_H^\dagger
    \end{pmatrix}.
\end{equation}
The reflection operator becomes 
\begin{equation}
    \hat{R}_{(a)}=\big(2\ketbra{+}{+} - \id\big)_{(a)} = \hat{X}_{(a)}
\end{equation}
and the walk operators can be constructed as
\begin{equation}
\hat{W}=
    \begin{pmatrix}
        0 & \hat{U}_H  \\ \hat{U}_H^\dagger & 0
    \end{pmatrix},\quad
\hat{V}=
    \begin{pmatrix}
        0 & \hat{U}_H^\dagger  \\ \hat{U}_H & 0
    \end{pmatrix}.
\end{equation}
The isomorphism between $\hat{W}$ and \eqref{eq:Y_arccosH} is now clear since it results in mapping the ancilla from $\hat{Z}$ to $\hat{Y}$.
The quantum walk now generates
\begin{equation}
    \bra{+}_{(a)} \underbrace{\dots\hat{V}\hat{W}\hat{V}\hat{W}}_{m\text{ walk operators}} \ket{+}_{(a)} = \hat{T}_m(\h).
\end{equation}
Finally, we notice that
\begin{equation}
    \hat{X}_{(a)}\hat{S}^\dagger \hat{X}_{(a)} = \hat{S}.
\end{equation}
Therefore, the iteration of  the select operator already generates the desirable walk:
\begin{equation}
    \bra{+}_{(a)} \hat{S}^m \ket{+}_{(a)}=\hat{T}_m(\h).
\end{equation}


We now want to utilise the decomposition in order to exploit it for the evaluation of the DOS via our KPM inspired algorithm. The KPM generally works better at the centre of the Hamiltonian spectrum, where exponentially many states reside. Close to the spectral edges it can become unstable, particularly at smaller system size, and hence less reliable~\cite{weisseKernelPolynomialMethod2006}. Exploiting the fact that exponentially many eigenstates are at the centre of the spectrum, one can expand the $\arccos(\h)$ around $E_k=0$:
\begin{equation}
\label{eq:arcc}
\begin{split}
    \arccos(\h)&=\frac{\pi}{2}-\sum_{k=0}^{\infty} \frac{(2k)!}{2^{2k}(k!)^2(2k+1)}\h^{2k+1}\\ &=\lim_{K\rightarrow \infty}\Bigg[\frac{\pi}{2}-\underbrace{\sum_{k=0}^{K}c_k\h^{2k+1}}_{\h_K}\Bigg].
\end{split}
\end{equation}
It follows that the select operator (now depending $K$) becomes
\begin{equation}
\label{eq:easy_select}
    \hat{S}_{K} = e^{i\hat{Z}\otimes(\pi/2-\h_K)}=e^{i\pi\hat{Z}/2}e^{-i\hat{Z}\otimes\h_K},
\end{equation}
and 
\begin{equation}
\label{eq:easy_select_walk}
    (\hat{S}_K)^m=e^{im\pi\hat{Z}/2}e^{-im\hat{Z}\otimes\h_K}.
\end{equation}
The select operator now can be implemented with just two controlled operators as
\begin{equation}
    \hat{S}_K = \ketbra{0}{0}\otimes \hat{U}_{H_{K}} + \ketbra{1}{1}\otimes \hat{U}_{H_{K}}^\dagger.
\end{equation}
In what follows we choose to implement the $\hat{U}_{H_K}$ using the standard  ST decomposition since this decomposition is already known to require adequate resources on the hardware platform~\cite{childsTowardFirstQuantumSimulation2018, childsTheoryTrotterError2021}. A key limitation of the ST formulae in general is that the only operator it can implement is the evolution operator. However, the combination of the arc-cosine approximation, ST decomposition and qubitization  allow us to approximate polynomials of $\h$ which is sufficient for our goal. We leave the explicit qubitization of $\hat{U}_H$ for future work.

From \eqref{eq:easy_select_walk} we see that the Chebyshev moments can be written as
\begin{equation}\label{eq:cheb_approx}
\begin{split}
    &T_{2m}\big(\h\big) \simeq (-1)^{m}\cos \big(m\h_K\big),\\
    &T_{2m+1}\big(\h\big) \simeq (-1)^{m}\sin \big(m\h_K\big).
\end{split}
\end{equation}
Hence, a Hadamard test on top of the random state will correctly implement this operation, as depicted in \figref{final_circuit}. Note that, as shown in \figref{final_circuit}, a final $\hat{Z}(-\beta)$ rotation on the control qubit is needed to implement the component of $\hat{H}_K$ that is proportional to the identity. For $K=0$ this rotation is simply
\begin{equation}
    \hat{Z}(-\beta) = e^{i(a/2b)\hat{Z}}, 
\end{equation}
i.e.~$\beta = -a/b$, where $a$ and $b$ are defined in~\eqref{eq:normilised_H}. Note that since the random state is used to simulate the maximally mixed state, the circuit of \figref{final_circuit} is a variant of a DQC1.

\section{Results on the Quantinuum H1-1 device}
\label{sec:results}
\subsection{Example Model to be simulated}

As a test model to implement our algorithm on quantum hardware we choose the non-integrable spin-$\frac12$ XYZ Heisenberg chain with a staggered interaction along the $Z$ direction:
\begin{align}\label{eq:xyz_zz}
    \h =& \sum_{i=0}^{L-1} J_x\hat{X}_{(i)}\hat{X}_{(i+1)}\hspace{-.05cm} +\hspace{-.05cm}J_y\hat{Y}_{(i)}\hat{Y}_{(i+1)}\hspace{-.05cm}+\hspace{-.05cm}J_z \hat{Z}_{(i)}\hat{Z}_{(i+1)} \notag \\ 
    &+\hspace{-.05cm}\sum_{i=0}^{L-1} (-1)^{i}\Lambda \hat{Z}_{(i)}\hat{Z}_{(i+1)}.
\end{align}
By choosing $J_z=\Lambda$, we can add a small advantage from the perspective of gate count, without adding any symmetry to the Hamiltonian. Indeed, the exponentiation of $\h$ (\eqref{eq:xyz_zz}) can be split into two non-commuting terms: interactions between even-odd spins and between odd-even spins. The latter, when $\Lambda = J_z$, have non-zero couplings only along the X and Y direction. Then, even-odd terms require a gate composition of \figref[b]{gate_dec}, while odd-even terms require the even shallower \figref[a]{gate_dec}.

\begin{figure}
\centering{\includegraphics[width=1.0\columnwidth]{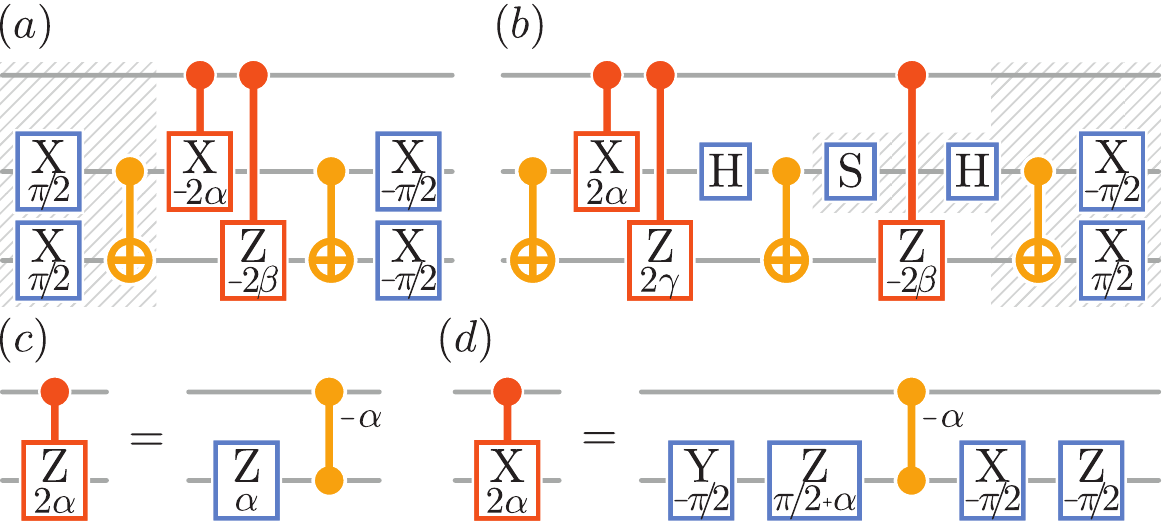}}
\caption{Gate decompositions used to implement the controlled unitary (i.e.~the orange gate of \figref{final_circuit}) for the Hamiltonian of \eqref{eq:xyz_zz}. The controlled ST decomposition of this Hamiltonian requires a controlled $\exp\big(i\alpha\hat{X}\otimes\hat{X}+i\beta\hat{Y}\otimes\hat{Y}\big)$ for the odd-even spin pairs (\textit{a}) and a controlled $\exp\big(i\alpha\hat{X}\otimes\hat{X}+i\beta\hat{Y}\otimes\hat{Y}+i\gamma\hat{Z}\otimes\hat{Z}\big)$ for the even-odd spin pairs (\textit{b}), with $\alpha,\beta,\gamma\in\mathbb{C}$. 
The hatched area in (\textit{a}) represents the part that for the first ST step will naturally mix with the last layer of the random circuit, without requiring additional gates; the hatched area of (\textit{b}) in the last ST step can be neglected.
The (\textit{c}) and (\textit{d}) respectively represent the conversion of the C-$\hat{Z}(2\alpha)$ and C-$\hat{X}(2\alpha)$ in the native gate set currently supported by the System Model H1-1. Note that this particular gate decomposition was chosen in order to require as few gates as possible on the control qubit, to enhance parallelisation and reduce the probability of errors occurring.}
\label{fig:gate_dec}
\end{figure}

\subsection{Hardware results}
\begin{figure}[!ht]
\centering{\includegraphics[width=1.0\columnwidth]{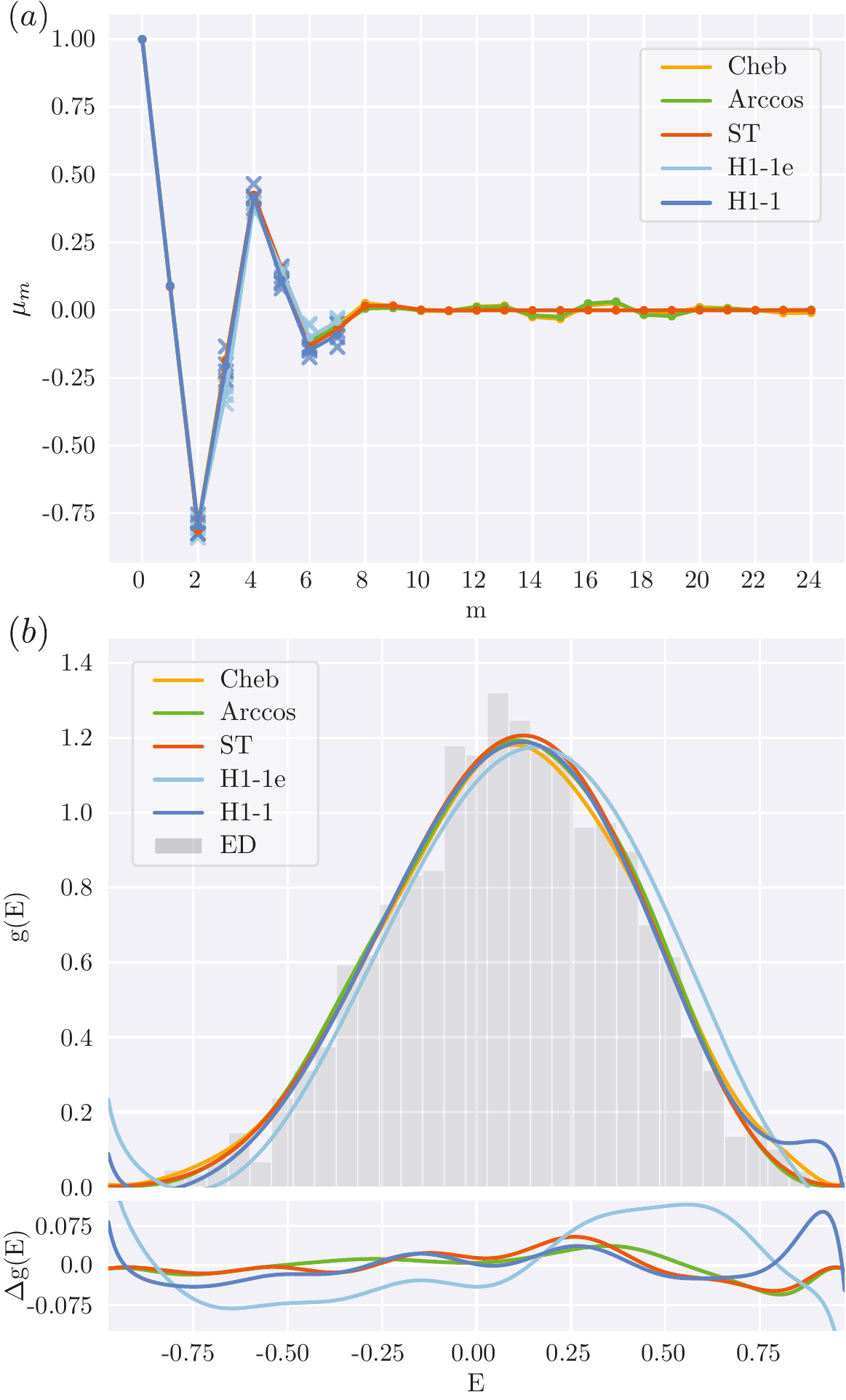}}
\caption{Results for the Hamiltonian of \eqref{eq:xyz_zz} at $L=12$ with $J_x=1,J_y=1/3,J_z=\Lambda=1/2$. In ($a$) the estimates for the first 25 Chebyshev moments. The comparison is between Chebyshev moments obtained in different ways: analytically (Cheb), applying the arc-cosine approximation at the first order (Arccos), adding the ST approximation with one single step (ST), and the results from the circuit of \figref{final_circuit} from Quantinuum, the emulator (H1-1e) and the quantum computer (H1-1). We used 4 different random circuits with $j=5$ and 1000 shots. Each cross represents the result from a circuit with a different random circuit. Figure ($b$) is composed by two parts. On top: the DOS obtained with the five different estimates for the Chebyshev moments (using a kernel with $M=25$) and the DOS from the exact diagonalisation (ED). On bottom we reported the comparison of the DOS obtained with different approximation of the Chebyshev moments w.r.t. the analytical ones.}
\label{fig:L12}
\end{figure}

\begin{figure}[!ht]
\centering{\includegraphics[width=1.0\columnwidth]{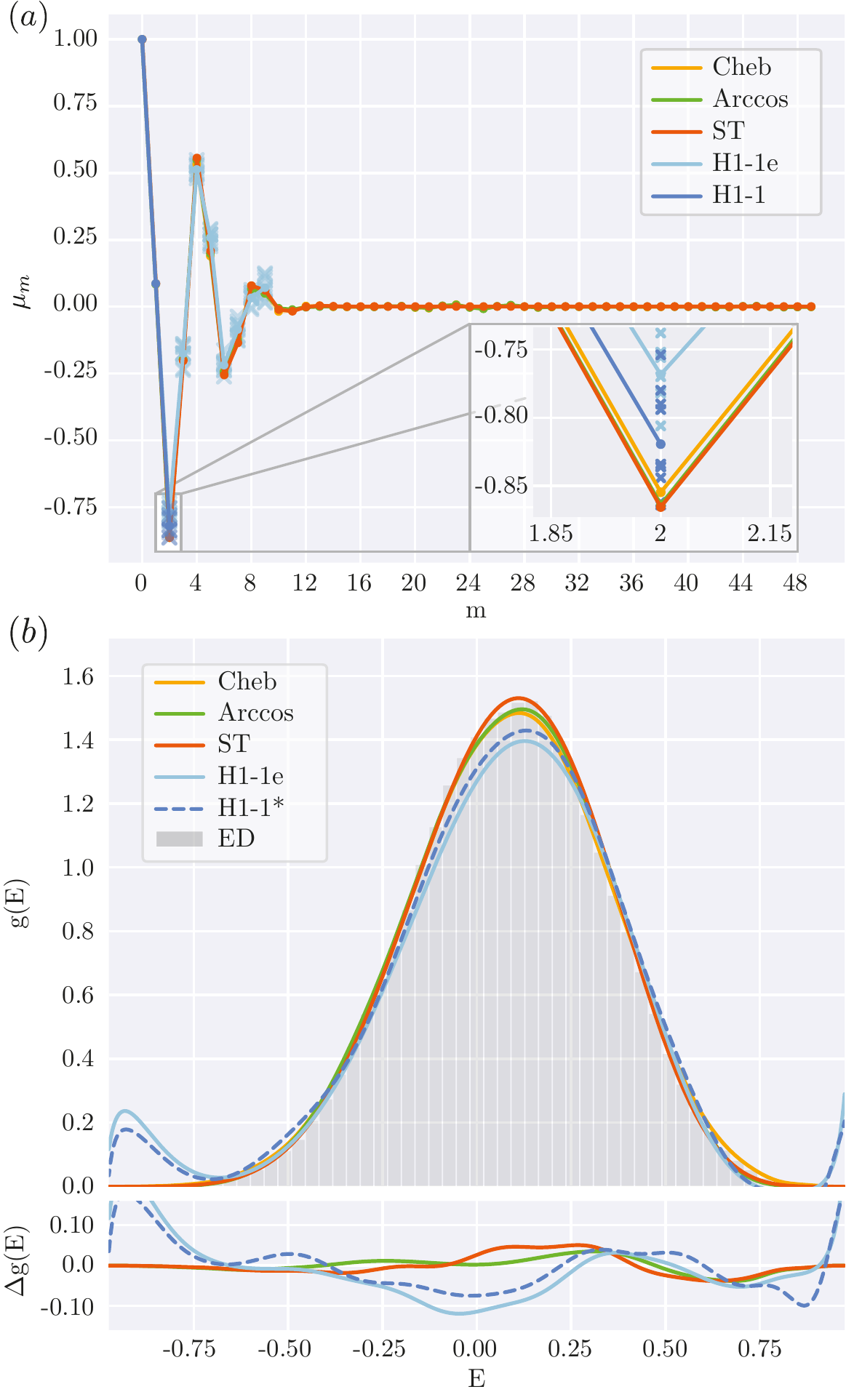}}
\caption{Similarly to \figref{L12}, the results for the Hamiltonian of \eqref{eq:xyz_zz} at $L=18$ with $J_x=1,J_y=1/3,J_z=\Lambda=1/2$. In ($a$) the estimates for the first 50 Chebyshev moments. Here, for the Quantinuum executions, we used 10 different random circuits with $j=4$ and 1000 shots. Each cross represents the result from a circuit with a different random circuit. Figure ($b$) is composed by two parts. On top: the DOS obtained with the five different estimates for the Chebyshev moments and the DOS from the exact diagonalisation (ED). Since we computed $\mu_2$ only from the H1-1 System, we also included what the DOS estimate from the $\mu_2$ from the real hardware combined with the estimates from the emulator would look like (H1-1*). On bottom we reported the comparison of the DOS obtained with different approximation of the Chebyshev moments w.r.t. the analytical ones. }
\label{fig:L18}
\end{figure}

We are now in a position to test our quantum algorithm on physical hardware using the model described above.  To evaluate the effectiveness of the method at each stage of approximation, we conducted quantum and classical simulations of the chain defined by \eqref{eq:xyz_zz} at $L=12$ and $L=18$. In \figref{L12}, we compare the Chebyshev moments obtained by various methods: by analytical calculation (using eigenvalues obtained from ED), by approximating the $\arccos$ function, by including the ST approximation, and by simulating the circuit of \figref{final_circuit} with and without noise. 
To generate the random state, we implemented a series of gates on the $L$-qubit register via the Par. method, as described in \secref{subsec:randomiser}. The Chebyshev moments were computed using the methods described in \secref{subsec:implenting_cheb}, taking the arccosine expansion to order $K=0$, i.e.~$\arccos(\hat{H})\approx \pi/2-\hat{H}$, and using a single ST step. Remarkably, we find that even with these parameters it is possible to achieve an appreciable precision. We found that hardware errors overcome any attainable improvements from using arccosine expansions with $K>0$ or more than a single ST step. More detail on the error analysis is given in \appref{err_analysis}. 

We found that for $L=12$, the KPM expansion reaches a sufficient degree of convergence to ED values at $M\sim 25$. Due to resource limitations on the Quantinuum device we used only four different random states, running 1000 shots for each one of these. We simulated up to $m=7$ on the quantum hardware, and approximate $\mu_m\approx 0$ for $m>7$, since these moments are too small to be distinguished from zero with our resources (see \appref{err_analysis}). \figref[b]{L12} shows that the moments extracted from the quantum computer are almost indistinguishable from the exact values within the bulk of the spectrum, and that the KPM with $M=25$ is able to accurately reconstruct the DOS using these moments. 

We next attempted the same calculation on a register of $L=18$, which uses $19$ out of $20$ qubits currently available on the H1-1 System~\cite{quantinuum}.
Here the KPM requires $M\sim 50$ to achieve an accurate approximation, while the moments that can be simulated accurately range up to $m=11$, beyond which their values become too small to be distinguished from zero given the number of shots we have used.
Simulating a larger system requires even more resources, and due to the higher costs associated with these larger circuits, we ultimately only simulated $\mu_2$ on the real hardware for the $L=18$ case. Our results demonstrate that the emulator (H1-1e) consistently produces accurate outcomes and we compute the rest of the moments with it. Additionally, we note that the fidelity of the $\hat{ZZ}$ gate appears to improve with smaller angles. Our circuit heavily employs these gates (see \figref{gate_dec}), and any discrepancies in the emulator results can be attributed to an underestimation of these gates' fidelities. We report the results from this last system in \figref{L18}, where in the plot of the DOS we add a projection obtained by combining the results from hardware and emulator together. 
 
\section{Conclusions}

In summary, we have performed the first estimation of the density of states of an non-integrable many-body quantum system on a digital quantum simulator. We have designed and implemented a quantum algorithm that exploits a combination of the Hadamard test, Suzuki-Trotter decomposition and random state preparation to extract Chebyshev moments. Proof-of-principle hardware simulations were performed on registers of $L=12$ and $L=18$ qubits on the Quantinuum H1-1 ion trap quantum computer, obtaining a good approximation to the DOS for a non-integrable Hamiltonian in the bulk of the spectrum (corresponding to high microcanonical temperatures). We explored in detail the crucial subroutines of stochastic trace evaluation and controlled evolution with arccosine approximation. We believe that our quantum hardware results represent the current state-of-the-art, in terms of both the generation of pseudo-random states and the implementation of controlled unitary operations on a many-qubit register. We emphasise that the accuracy of our hardware results has been limited primarily by financial constraints, and not by fundamental resource scalings nor even by noise on the H1-1 device.

For the DOS we found it was sufficient to take the arccosine expansion to very low order ($K=0$), which is ultimately due to the concentration of energy levels at the centre of the spectrum in large systems. We note that our KPM-inspired approach can easily be tailored to compute finite-temperature expectation values in the diagonal and micro-canonical ensembles, in addition to other spectral functions such as the Lehmann representation of multi-time correlation functions. In these cases, it may be necessary to consider higher-order expansions (i.e.~$K>0$) to account for features away from the centre of the spectrum. Our methods could also be combined with other quantum algorithms tailored to compute ground-state and low-lying excited state properties~\cite{kassal2011simulating,hastings2015improving,cao2019quantum,de2021materials,cerezo2021variational,RevModPhys.94.015004}, in order to estimate thermodynamic properties across the full range of temperature scales.

Our estimation of the DOS on current quantum hardware represents an important step forward towards quantum statistical mechanics calculations on quantum computers. As the devices improve, we expect that this algorithm and subroutines can be used to extract useful approximations to thermodynamic properties in regimes not accessible to state-of-the-art classical numerical techniques for strongly correlated systems. 

\section{Acknowledgements} 
 JG would like to thank Sabrina Maniscalco for inviting him to take part in an unconference in Lapland where some of the first discussions related to this work took place. We thank Microsoft Ireland, in particular Kieran McCorry, for providing generous funding to run this project and and for providing access to the Quantinuum machine through Microsoft Azure Quantum. JG is supported by a SFI-Royal Society University Research Fellowship and acknowledges funding from European Research Council Starting Grant ODYSSEY (Grant Agreement No. 758403). MTM is supported by a Royal Society-Science Foundation Ireland University Research Fellowship (URF\textbackslash R1\textbackslash 221571), and acknowledges funding from the European Commission via the Horizon Europe project ASPECTS (Grant Agreement No. 101080167).

\appendix
\section{Chebyshev polynomials}
\label{app:chebyshev_polynomials}
Given a $f$ function, $\{g_M\}$, a family of orthogonal functions s.t.
\begin{equation}
    f(x)\approx g_M(x) =\sum_{m=0}^M a_m \Phi_m(x)
\end{equation}
is said to be a good approximation if it approximates $f$ in at least the square norm:
\begin{equation}
    \norm{f(x)-g_M(x)}_2=\left[\int\left(f(x)-g_M(x)\right)^2dx\right]^{1/2}.
\end{equation}

A family of functions that are frequently used for this scope is the given by Fourier series. In this case, the polynomials are:
\begin{equation}
    g_M(x)=\frac12 a_0 + \sum_{m=1}^M\left(a_m\cos(m x) + b_m \sin(m x)\right)
\end{equation}
where the coefficients are 
\begin{equation}\label{eq:fourier_coeff}
    \begin{split}
        a_m &= \frac{1}{\pi} \int f(x) \cos(mx) dx\\
        b_m &= \frac{1}{\pi} \int f(x) \sin(mx) dx
    \end{split}
\end{equation}
The Fourier decomposition works well with signals, namely with processes that happen to be periodic and extended in time (its convergence domain is an infinite strip, symmetric around the real axes~\cite{boydChebyshevFourierSpectral}). 
Similarly, we can evaluate $f$ (and the coefficients $a_m$ and $b_m$) just in the $[-\pi, \pi]$ interval.
The Fourier series is then \textit{exponentially convergent} for periodic functions with derivatives bounded in $[-\pi, \pi]$, namely its coefficients $a_m$ and $b_m$ decrease exponentially in $m$:
\begin{equation}
    a_m, b_m \sim \mathcal{O}[ \exp (-qn^r)]
\end{equation}
with $n\gg1$ and $q$ a constant for some $r>0$. If $f$ is even(odd) all the sine(cosine) coefficients will cancel out and the expansion becomes the \textit{Fourier sine(cosine) series}.\\

Many physical phenomena are not periodic but bounded (occur in a limited space). 
So, to make a function behave periodically one can apply a change of variable
$$x= \cos(\theta)$$
with then $\theta\in [-\pi, \pi]$ $\rightarrow$ $x\in [-1, 1]$.
Noticing that $f(\cos(\theta))=f(\cos(-\theta))$ the expansion reduces to the Fourier cosine series:
\begin{equation}
    g_M(\cos(\theta) ) = \sum_{m=0}^Ma_m \cos(m\cos(\theta))
\end{equation}
where $\cos(m\cos(\theta))$ are the Chebychev polynomials $T_m(\cos(\theta))$. Reformulating this last equation in $x$ we have
\begin{equation}
    g_M(x) = \sum_{m=0}^M a_m T_m(x)
\end{equation}
where now 
\begin{equation}\tag{\ref{eq:cheb_as_cosarc}}
    T_m(x)=\cos(m\arccos(x)).
\end{equation}
Due to this strong connection with the Fourier series, the Chebyshev series inherits all its properties (among which the exponential convergence) with now the advantage to be working also with non-periodic functions. \\

We can notice that $\cos(m\arccos(x))$ is actually a polynomial in $x$ by looking at the identity 
\begin{equation}
    \cos(2\phi) = \cos^2(\phi) -1  
\end{equation}
which leads to the following iterative property of the Chebychev polynomials:
\begin{equation}\label{eq:recursive_cheb}
    \begin{split}
        T_0(x) &= 1,\\
        T_1(x) &= x,\\
        T_{m+1}(x) &= 2x T_m(x) - T_{m-1}(x).
    \end{split}
\end{equation}

\section{Higher orders of the arc-cosine expansion}
\label{app:higher_order_expansion}

\begin{figure}[!ht]
\centering{\includegraphics[width=1.\columnwidth]{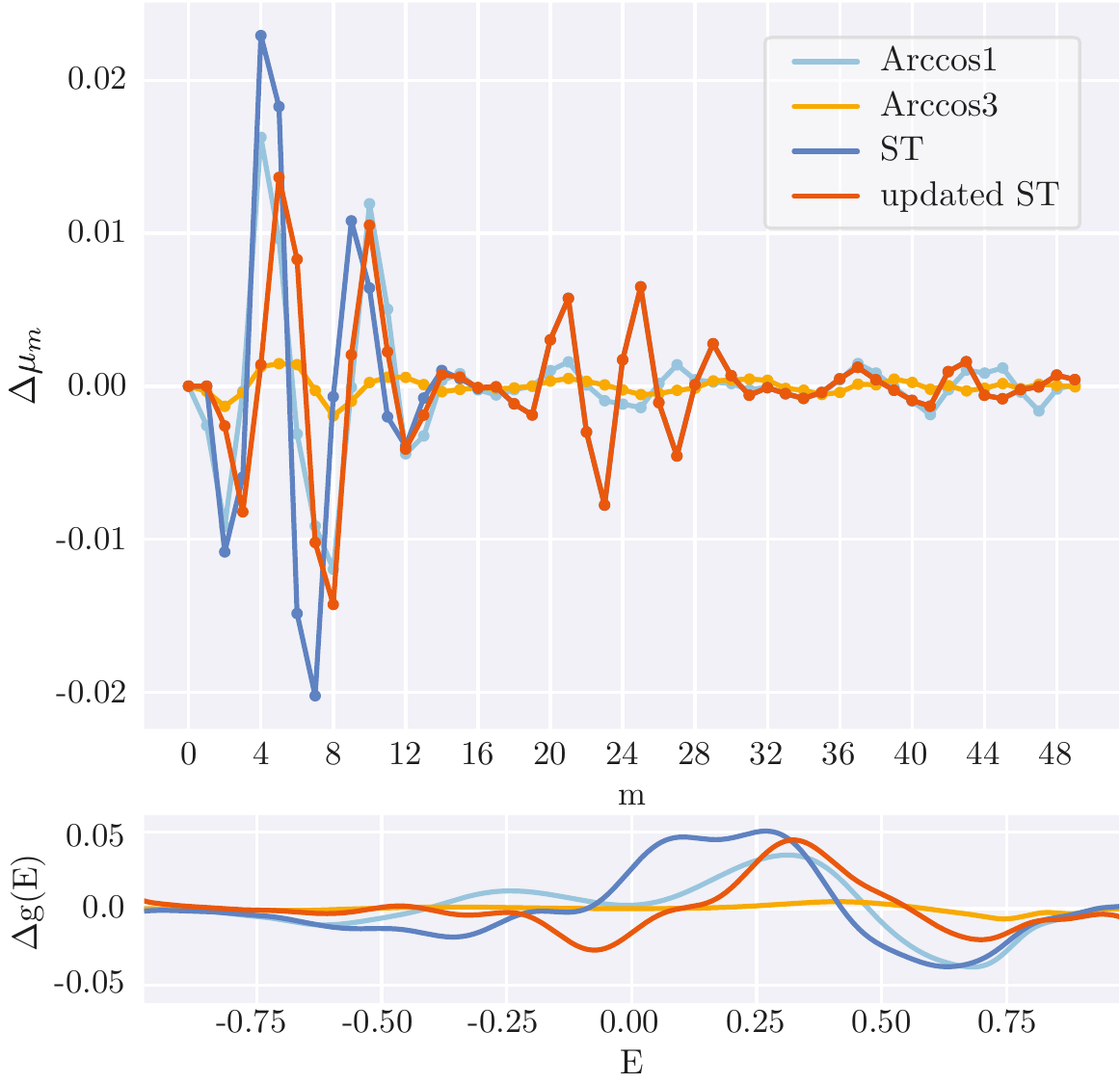}}
\vspace{-.75cm}
\caption{Comparison of the first 50 analytical Chebyshev moments with different approximations: the first order arc-cosine approximation (Arccos1), the second, which involves the $3^\text{rd}$ power of the Hamiltonian (Arccos3), the normal ST approximation at the first order (ST) and, its version with the \textit{updated} parameters (updated ST). After $m=15$ the difference between these last two approximations becomes negligible. }
\label{fig:updated_parameters}
\vspace{-.25cm}
\end{figure}
Due to the concentration of eigenstates around zero, in a lot of physical Hamiltonians, the first-order expansion is already enough to shape the DOS with a good approximation. However, in case a better precision is required one can truncate the $\arccos$ approximation to the second order (i.e. $K=1$). At this order of approximation, a $k$-local Hamiltonian would become $3k$-local. However, the terms that are actually $3k$-local contribute to a lesser degree.
Moreover, a large part of $\h^3$ will lie on the same operators of $\h$.
For instance, in the Hamiltonian of \eqref{eq:xyz_zz} the only coefficients in $\h^3$ that increase with the system size are the ones of Pauli operators already present in $\h$.
Indeed, one can simply implement the same combination of Pauli operators of $\h$ by \textit{updating} the parameters to include the contributions from $\h^3/6$.
These updated parameters will be, considering the case $J_z+\Lambda=J_x$, for the $J_x$ of the even sites
\begin{equation}
J_x\bigg(1+\frac16 \Big((9L/2-2)J_x^2 + (3L-4)J_y^2 + 3\beta^2 + 6J_y\beta\Big)\bigg),
\end{equation}
for the $J_x$ of the odd sites
\begin{equation}
J_x\bigg(1+\frac16 \Big((9L/2-6)J_x^2 + (3L-4)J_y^2 + 3\beta^2\Big)\bigg).
\end{equation}
Similarly for $J_y$ of even sites:
\begin{equation}
    J_y+\dfrac16\Big((3L-2)J_y^3 + (9L/2-4)J_yJ_x^2 + 3J_y\beta^2 + 6J_x^2\beta\Big)
\end{equation}
and for the odd ones:
\begin{equation}
    J_y\bigg(1+\dfrac16\Big((3L-2)J_y^2 + (9L/2-8)J_x^2 + 3\beta^2\Big)\bigg).
\end{equation}
Then, for $z$-coupling our Hamiltonian only has even terms with a coefficient updated as:
\begin{equation}
    J_x\bigg(1+\dfrac16\Big(((9L/2-6)J_x^2 + (3L-4)J_y^2 + 3\beta^2 + 6J_y\beta)\Big)\bigg)
\end{equation}
and finally, the coefficient of the identity becomes:
\begin{equation}
    \beta +\dfrac16\Big(-\beta^3 - 3L J_y^2 - 9/2L\beta J_x^2 - 3LJ_x^2J_y\Big).
\end{equation}
The comparison of how the ST decomposition is affected by using these parameters instead of the initial one is reported in \figref{updated_parameters}.\\


\section{Powers of the Hamiltonian}
\label{app:powers_of_h}
In the appendix, we see how a similar approach to the arc-cosine approximation can be adapted to implement the powers of the Hamiltonian. In~\cite{kazuhiroQuantumPowerMethod2021} Seki and Yunoki devise a method to implement powers of a Hamiltonian as the finite time derivative of the evolution operator, based on the idea that
\begin{equation}
    \h^m = i^m \frac{\mathrm{d}^m \hat{U}(t)}{\mathrm{d}t^m} \bigg\lvert_{t=0}.
\end{equation}
At a finite time $t$, the authors suggest taking the central derivative and for the first power of the Hamiltonian the equation becomes
\begin{equation}
    \h\simeq\frac{i}{2t}\left(e^{-it\h}-e^{it\h}\right).
\end{equation}
This is equivalent to approximating the Hamiltonian as
\begin{equation}
    \h\simeq \frac{e^{it\h}-e^{-it\h}}{2it} = \frac{\sin(t \h) }{t}.
\end{equation}
A closer look reveals that the first order approximation of the sine corresponds to the first order approximation of the arc-sine in
\begin{equation}
\begin{split}
    \h &= \frac{e^{i\arcsin{t\h}}-e^{-i\arcsin{t\h}}}{2it}\\
    &\simeq \frac{1}{2it}\left(e^{i(t\h)_K}-e^{-i(t\h)_K}\hspace{-.075cm}\right).
\end{split}
\end{equation}
Namely an alternative two-unitary decomposition to the one of \eqref{eq:simple_decomp_of_h} with the arc-sine instead of the arc-cosine in \eqref{eq:u_arccosH}.
This offers a path to increasing the precision of the method in a more controlled way. 
Furthermore, the coefficients of the expansion can be tailored so that the powers of $\h$ in the expansion will elide together:
\begin{equation}
    c_k=-\dfrac{(-1)^{k}}{(2k+1)!}-\frac{(-1)^{k'}(c_{k'})^{2k'+1}}{(2k'+1)!}
\vspace{.1cm}
\end{equation}
if $\exists\ k'\in\mathbb{N}$ s.t. $2k'+1=\sqrt{2k+1}$, and
\begin{equation}
    c_k=-\frac{(-1)^{k}}{(2k+1)!}-\frac{(-1)^{n'}c_{k'}^{2n'+1}}{(2n'+1)!}-\frac{(-1)^{k'}c_{n'}^{2k'+1}}{(2k'+1)!}
\end{equation}
if $\exists\ k',n'\in\mathbb{N}, k'\neq n'$ s.t. $2k+1=(2k'+1)(2n'+1)$, and otherwise
\begin{equation}
    c_k=-\frac{(-1)^{k}}{(2k+1)!}.
\end{equation}
In this way the error will scale as $\mathcal{O}((t\h)^{2K+1})$.\\

As highlighted, the probability of successfully selecting the combination of unitaries will depend on the magnitude of the trace of the product of $t$ and $\h$. Therefore, if the Hamiltonian has a large trace, as 
\begin{equation}
    \h=J\sum_{i} \id + \hat{X}_{(i)}\hat{X}_{(i+1)} + \hat{Y}_{(i)}\hat{Y}_{(i+1)} + \hat{Z}_{(i)}\hat{Z}_{(i+1)}
\end{equation}
it will be possible to take small time steps and still have a non-negligible trace. Otherwise, if the Hamiltonian is normalised it will necessary to take $t\lesssim 1$.

\section{Error analysis}
\label{app:err_analysis}
The proposed method seeks to achieve efficient execution on NISQ devices by prioritizing the creation of shallow, noise-resilient circuits over high precision. This is done while adhering to the constraint of limited time on the QPUs~\footnote{From this last, it is necessary to prioritize the requirement of finding good (so that a few are enough) and shallow random circuits rather than relying on MPS, made up of single layers of single rotation gates.  In fact, the cost of the circuit is dictated by the equation~\cite{quantinuum} 
\begin{equation}
    \text{HQC}=5+\frac{\text{shots}}{5000}(N_{1q}+10N_{2q}+5N_m)
\end{equation} 
where $N_{1q}, N_{2q}$ are the number of single and double qubit gates, respectively, while $N_m$ is the number of qubits measured}. 
We have to take into account several sources of error in comparison to the analytical Chebyshev moment $m$.\\
The first is due to the arc-cosine approximation of \secref{subsec:implenting_cheb}. In this case, the error depends on the distribution of the eigenvalues of the Hamiltonian. For odd polynomials, the total expansion will be
\begin{equation}
\begin{split}
    &\ \hat{T}_{2m+1}(\h)=(-1)^m\sin((2m+1)\h_{K\rightarrow \infty})=\\
    &=(-1)^m\sum_{n,k} \frac{(-1)^n}{(2n+1)!}\left[\frac{(2m+1)(2k)!}{2^{2k}(k!)^2(2k+1)}\h^{2k+1}\right]^{2n+1}
\end{split}
\end{equation}
and similarly for the even terms 
\begin{equation}
\begin{split}
    &\ \hat{T}_{2m}(\h)=(-1)^m\cos((2m)\h_{K\rightarrow \infty})=\\
    &=(-1)^m\sum_{n,k} \frac{(-1)^n}{(2n)!}\left[\frac{(2m)(2k)!}{2^{2k}(k!)^2(2k+1)}\h^{2k+1}\right]^{2n}.
\end{split}
\end{equation}
The first contributes to the error are 
\begin{equation}
\label{eq:main_contributes@K=0}
    \varepsilon_{o}\simeq\frac{2m+1}{6}\tr[\h^{2K+3}]\ \text{ and }\ \varepsilon_{e}\simeq\frac{m^2}{18}\tr[\h^{4K+6}]
\end{equation}
for even and odd moments respectively.
To roughly characterize it, we need to introduce speculation about the spectrum of the system. In many physical systems, the DOS can be modelled as a Gaussian distribution~\cite{hartmannSpectralDensitiesPartitionFunctions2005}, idea behind which seems also to be inspired the GIT method~\cite{roggeroSpectralDensityEstimation2020}.
If the variance of the DOS is $\Delta_E$ and $\bar{E}$ the average, the Gaussian curve that describes the spectrum of $\h$ is
\begin{equation}
    g(E)\simeq\frac{1}{\sqrt{2\pi}\Delta_E}\exp\left(-\frac{(E-\bar{E})^2}{2\Delta_E^2}\right).
\end{equation}
For $K=0$, the contributes of  \eqref{eq:main_contributes@K=0} become
\begin{align}
    \varepsilon_{o}=&\frac{2m+1}{6}\int E^3g(E)\mathrm{d}E\notag \\ \simeq & \frac{2m+1}{6}\bar{E}(\bar{E}^2+3\Delta_E^2), 
\end{align}
\begin{align}
    \varepsilon_{e}=&\frac{m^2}{18}\int E^6g(E)\mathrm{d}E\simeq \notag \\
    \simeq&\frac{m^2}{18}(\bar{E}^6+15\Delta_E^2\bar{E}^4+45\Delta_E^4\bar{E}^2+15\Delta_E^6).
\end{align}
Note that following the normalisation of \eqref{eq:normilised_H}, both $\bar{E}$ and $\Delta_E$ will be less than 1. Choosing $\Delta_E\geq \bar{E}$ (as it is in our case) justifies why in our results we did not see better performances at even rather than odd moments.

The \textit{second} to the ST decomposition, to decompose the operator in simple gates, if the ST order is the first and the amount of steps is $1$ the error will go as $\mathcal{O}(5^t m^{1+\frac{1}{t}} /\varepsilon^\frac{1}{t})$ where $t$ is the ST order~\cite{suzukiGeneralTheoryFractalPath1991}.\\
The \textit{third} follows from the stochastic trace estimation of \secref{sec:stoch_trace}, where the error is $\mathcal{O}\big(1/\sqrt{R 2^L}\big)$.\\
The \textit{fourth} is due to the circuit and the finiteness of the number of shots: 
\begin{equation}
    \sqrt{\frac{1-\mu_m}{R\ \text{shots}}}\sim\mathcal{O}\left(\frac{1}{\sqrt{R\ \text{shots}}}\right).
\end{equation}

\bibliographystyle{apsrev4-1}
\bibliography{bibl}

\end{document}